\input amstex
\magnification=1200
\loadbold
\loadeufm
\loadmsbm
\vsize=18.true cm
\voffset=-1truecm
\hsize=13.5true cm

\TagsOnRight

\def\frf#1{\vcenter{\hrule\hbox{\vrule\kern1pt
\vbox{\kern1pt\hbox{$\displaystyle#1$}%
\kern1pt}\kern1pt\vrule}\hrule}}
\def\notni{\hbox{{\kern0pt\raise.7pt\hbox{${\scriptstyle +}$}}
{\kern-9.5pt\raise0pt\hbox{$\supset$}}}}

\centerline{\bf IRREDUCIBLE REPRESENTATIONS OF CAYLEY--KLEIN}

\medskip
\centerline{\bf ORTHOGONAL ALGEBRAS}

\bigskip
\centerline{\bf N.A.~Gromov}

\medskip
\centerline{Mathematical Department of Komi Research Centre}

\centerline{of Russian Academy of Sciences,}

\centerline{Kommunisticheskaja str. 24, Syktyvkar}

\centerline{Komi Republic, Russia}

\centerline{E-mail: gromov\@omkomi.intec.ru}

\bigskip
\centerline{\bf S.S.~Moskaliuk}

\medskip
\centerline{Bogolyubov Institute for Theoretical Physics}

\centerline{of National Academy of Sciences of Ukraine}

\centerline{Metrolohichna str. 14-b,}

\centerline{252143, Kyyv-143, Ukraine}

\centerline{E-mail: mss\@bitp.kiev.ua}

\vskip2cm
 \centerline{\bf Abstract}

\medskip
 Multidimensional contractions of irreducible representations
of the Cayley--Klein orthogonal algebras in Gel'fand--Zetlin
 basis are considered. Contracted over different parameters, algebras
 can turn out to be isomorphic. In this case method of transitions
 describes the same reducible representations in different basises,
 say discrete and continuons ones.

\vfill\eject
{\bf 1. Algebra $so\,(3;\boldkey j)$}
\medskip

In spite that algebra $so\,(3)$ is isomorphic to algebra $u\,(2)$, we
shall consider it separately, because algebras $so\,(3;\boldkey j)$,
$\boldkey j=(j_1,j_2)$, in contrast to $u\,(2;j_1)$, allow contraction
over two parameters. Under transition from $so\,(3)$ to
$so\,(3;\boldkey j)$ generators are transformed as follows:
$X_{01}=j_1X^*_{01}(\to)$, $X_{02}=j_1j_2X^*_{02}(\to)$,
$X_{12}=j_2X^*_{12}(\to)$,
and the only Casimir operator is transformed as
$C_2(\boldkey j)=j_1^2j_2^2C^*_2(\to)$ [1,2].
Gel'fand and Zetlin
[3] have also found irreducible representations of algebra
$so\,(3)$ in the basis, determined by the chain of subalgebras
$so\,(3)$ $\supset$ $so\,(2)$; they are
given by operators
$$\gathered
X^*_{12}|m^*\rangle=im^*_{11}|m^*\rangle,\\
X^*_{01}|m^*\rangle={1\over2}\bigl\{\sqrt{(m_{12}^*-m_{11}^*)(m_{12}^*+m_{11}^*
+1)}|m_{11}^*+1\rangle-\bigr.\\
-\sqrt{(m_{12}^*+m_{11}^*)(m_{12}^*-m_{11}^*+1)}|m_{11}^*
-1\rangle\bigl.\bigr\}, \\
X^*_{02}|m^*\rangle={i\over2}\bigl\{\sqrt{(m_{12}^*-m_{11}^*)(m_{12}^*+m_{11}^*
+1)}|m_{11}^*+1\rangle-\bigr.\\
-\sqrt{(m_{12}^*+m_{11}^*)(m_{12}^*-m_{11}^*+1)}|m_{11}^*
-1\rangle\bigl.\bigr\},\endgathered\tag1$$
where the schemes $|\,m^*\rangle$, enumerating the elements of
Gel'fand-Zetlin orthonormed basis, are
$|\,m^*\rangle=\left|\left.\matrix
m^*_{12}\\m^*_{11}\endmatrix\right.\right>$, $|m^*_{11}|\leq m^*_{12}$
and components $m^*_{11}$, $m^*_{12}$ are simultaneously either
integer, or half-integer. Spectrum of Casimir operator is
$C^*_2(m^*_{12})=m^*_{12}(m^*_{12}+1)$.

Component $m^*_{11}$ is eigenvalue of operator $X^*_{12}$, for this
reason its transformation is determined by transformation $X^*_{12}$,
i.e. $m_{11}=j_2m^*_{11}$. The transformation of component $m^*_{12}$
can be found from the requirement of determinacy and non-zero spectrum
of Casimir operator $C_2(m_{12})=\mathbreak=j_1^2j_2^2C^*_2(\to)
=j_1^2j_2^2\dfrac{m_{12}} {j_1j_2}
\biggl(\dfrac{m_{12}}{j_1j_2}+1\biggr)=m_{12}(m_{12}+j_1j_2)$ under
contractions, i.e. $m_{12}=j_1j_2m^*_{12}$. Then we find
from (1) operators of representation of algebra $so\,(3;\boldkey
j)$
$$
\gather X_{12}|m\rangle=im_{11}|m\rangle,\\
X_{01}|m\rangle={1\over2j_2}\bigl\{\sqrt{(m_{12}-j_1m_{11})(m_{12}+j_1m_{11}
+j_1j_2)}|m_{11}+j_2\rangle-\bigr.\\
-\sqrt{(m_{12}+j_1m_{11})(m_{12}-j_1m_{11}+j_1j_2)}|m_{11}
-j_2\rangle\bigl.\bigr\},\tag2\\
X_{02}|m\rangle={i\over2}\bigl\{\sqrt{(m_{12}-j_1m_{11})(m_{12}+j_1m_{11}
+j_1j_2)}|m_{11}+j_2\rangle+\bigr.\\
+\sqrt{(m_{12}+j_1m_{11})(m_{12}-j_1m_{11}+j_1j_2)}|m_{11}
-j_2\rangle\bigl.\bigr\},\endgather$$
where the components of scheme $|\,m\rangle$ satisfy inequalities
$|m_{11}|\leq m_{12}/j_1$. Operators (2) satisfy commutation
relations  of algebra $so\,(3;\boldkey j)$, which can be checked
up straightforwardly [1]. The representation is irreducible. This can
be established, as in the case of unitary algebras, by action of
rising and lowering operators on vectors of major and minor weights.

For $j_1=\iota_1$ the relations (2) give irreducible representation
of algebra $so\,(3;\iota_1,j_2)\equiv IO(2;j_2)=\{X_{01},X_{02}\}$ $\notni$
$\{X_{12}\}$:
$$\gather
X_{12}|m\rangle=im_{11}|m\rangle,\\
X_{01}|m\rangle={m_{12}\over2j_2}(|m_{11}+j_2\rangle-|m_{11}-j_2\rangle),
\tag3\\
X_{02}|m\rangle={i\over2}m_{12}(|m_{11}+j_2\rangle+|m_{11}-j_2\rangle),
\endgather$$
where $m_{11}$ integer or half-integer, and $m_{12}\in\Bbb R$,
$m_{12}\geq0$. The eigenvalues of Casimir operator
$C_2(\iota_1,j_2)$ are $m_{12}^2$.

For $j_2=\iota_2$ we obtain from (2) irreducible representation of
algebra $so\,(3;j_1,\iota_2)=\{X_{02},X_{12}\}$ $\notni$ $\{X_{01}\}$:
$$
\gathered
X_{12}|m\rangle=im_{11}|m\rangle,\quad
X_{02}|m\rangle=i\sqrt{m_{12}^2-j_1^2m_{11}^2}|m\rangle,\\
X_{01}|m\rangle=\sqrt{m_{12}^2-j_1^2m_{11}^2}|m\rangle'_{11}-j_1^2{
m_{11}\over2\sqrt{m_{12}^2-j_1^2m_{11}^2}}|m\rangle,
\endgathered
\tag4
$$
where $m_{11},m_{12}\in\Bbb R$ and $|m_{11}|\leq m_{12}$.

Ceasing to fix the coordinates $x_0,x_1,x_2$ in $\Bbb R_3(\boldkey
j)$, where group $SO(3;\allowmathbreak\boldkey j)$ acts, we can easily
prove isomorphism of algebras $so\,(3;\iota_1,1)$ and
$so\,(3;\allowmathbreak1,\iota_2)$.  Then (3) for $j_2=1$ and
(4) for $j_1=1$ give description of irreducible representation of
algebra $IO(2)$ in discrete and continuous basises, correspondingly,
in infinite-dimensional space of representation. Discrete basis
consists of eigenvectors of compact operator $X_{12}$ with integer or
half-integer eigenvalues $-\infty<m_{11}<\infty$. Continuous basis
consists of generalized eigenvectors of noncompact operator $X_{12}$,
which eigenvalues are $m_{11}\in\Bbb R$, $|m_{11}|\leq m_{12}$.
Celeghini [4--6] has considered close to ours approach to
contraction and irreducible representations of orthogonal groups
$SO(3)$, $SO(5)$, related with singular transformation of components
of Gel'fand-Zetlin schemes.

Two-dimensional contraction $j_1=\iota_1$, $j_2=\iota_2$ gives
irreducible representation of ( two-dimensional) Galilean group
(algebra) $so\,(3;\boldsymbol\iota)$:
$$
\gather
X_{01}|m\rangle=m_{12}|m\rangle'_{11},\quad
X_{12}|m\rangle=im_{11}|m\rangle,\\
X_{02}|m\rangle=im_{12}|m\rangle,\tag5
\endgather
$$
where $m_{12},m_{11}\in\Bbb R$, $m_{12}\geq0$,
$-\infty<m_{11}<\infty$;
$C_2{(\boldsymbol\iota)}=m_{12}^2$.

The result of action of generators on the derivative
$|\,m\rangle'_{11}\equiv\dfrac{\partial}{\partial
m_{\scriptscriptstyle 11}}|\,m\rangle$ can be found, applying a
generator to both sides of equation
$|\,m\rangle'_{11}=\mathbreak=\dfrac{2}{2\iota_2}(|\,m_{11}+\iota_2\rangle-
|\,m_{11}-\iota_2\rangle)$. In particular,
$X_{12}|\,m\rangle'_{11}=im_{11}\,|\,m\rangle'_{11}+\allowmathbreak+
i\,|\,m\rangle$.
\bigskip

{\bf 2. Algebra $so\,(4;\boldkey j)$}
\medskip

To determine an irreducible representation of algebra
$so\,(4;\boldkey j)$, $\boldkey j=\mathbreak=(j_1,j_2,j_3)$, it is
sufficient to give representation of generators
$X_{01},X_{12},\allowmathbreak X_{23}$. Using the formulas from
monograph [7], we can change indices of generators according to the
rule: $4\to0$, $3\to1$, $2\to2$, $1\to3$. Then Gel'fand- Zetlin
representation corresponds to the chain of subalgebras
$so\,(4;\boldkey j)$ $\supset$ $so\,(3;j_2,j_3)$
$\supset$ $so\,(2;j_3)$, where $so\,(4;\boldkey
j)=\{X_{\mu\nu},\ \mu<\nu,\ \mu,\nu=\allowmathbreak=0,1,2,3\}$;
$so\,(3;j_2,j_3)=\{X_{\mu\nu},\ \mu<\nu,\ \mu,\nu=1,2,3\}$;
$so\,(2;j_3)=\mathbreak=\{X_{23}\}$.  Representation of generators
$X_{23}$, $X_{12}$ is given by (2), where indices of generators and
parameters $\boldkey j$ must be increased by 1, and scheme
$|\,m\rangle=\allowmathbreak\left|\left.\matrix
m_{12}\\m_{11}\endmatrix\right.\right>$ has to be substituted for
$$
|\,m\rangle=\left|\matrix m_{13}\ \ \ m_{23}\\ m_{12}\\
m_{11}\endmatrix\right>.
\tag6
$$

The rule of transformation for components $m^*_{12}$, $m^*_{11}$ can
be found by consideration of algebra $so\,(3;\boldkey j)$:
$m^*_{11}=j_3m^*_{11}$, $m_{12}=j_2j_3m^*_{23}$, $|m_{11}|\leq
\allowmathbreak\leq m_{12}/j_2$. It remains to derive the
transformation of components $m^*_{13}$, $m^*_{23}$, which determine
irreducible representation. To this aim we consider spectrum of
Casimir operators for algebra $so\,(4)$, found by A.M.Perelomov and
V.S.Popov [8], A.N.Leznov, I.A.Malkin, V.I.Man'ko [9]:
$$C^*_2=m^*_{13}(m^*_{13}+2)+m^{*2}_{23},\quad
C^{*'}_2=-(m^*_{13}+1)m^{*}_{23},\tag7$$
as well as the rule of transformation
$$
\gathered
C_{2p}(\boldkey j;X_{\mu\nu})=\biggl(\prod_{m=1}^{p-1}
j_{m}^{2m}j_{n-m+1}^{2m}
\prod_{l=p}^{n-p+1}j_{l}^{2p}\biggr)C^*_{2p}
\biggl(X_{\mu\nu}\prod_{l=\mu+1}^\nu j_l^{-1}\biggr),\\
C_{n}'(\boldkey j;X_{\mu\nu})=
\biggl(j_{{n+1\over 2}}^{{n+1\over 2}}
\prod_{m=1}^{(n-1)/2}j_m^m j_{n-m+1}^m\biggr)
C^{*'}_n\biggl(X_{\mu\nu}\prod_{l=\mu+1}^\nu j_l^{-1}\biggr).
\endgathered
\tag8a
$$
for Casimir operators under transition from $so\,(4)$
to $so\,(4;\boldkey j)$ [1]:
$$
C_2(\boldkey j)=j_1^2j_2^2j_3^2C_2^*(\to),\quad
C'_2(\boldkey j)=j_1j_2^2j_3C_2^{*'}(\to).
\tag8
$$
Requiring eigenvalues of operators $C_2(\boldkey j)$ and
$C'_2(\boldkey j)$ to be determinate expressions under contractions,
we get from (7), (8) for $C'_2(\boldkey j)$
$$
m_{12}m_{23}=j_1j_2^2j_3m_{13}^*m_{23}^*.
\tag9
$$
This equation (if
transformations of components $m_{13}$, $m_{23}$ involve only the
first powers of parameters $\boldkey j$) gives possible rules of
transformation of these components.

Let us write down possible variants of transformations of irreducible
representations of algebra $so\,(4)$ into representations of algebra
$so\,(4;\boldkey j)$ as well as transformed spectra of Casimir
operators
$$\gather
1)\quad m_{13}=j_1j_2m_{13}^*,\quad m_{23}=j_2j_3m_{23}^*,\\
C_2(\boldkey j)=j_3^2m_{13}(m_{13}+2j_1j_2)+j_1^2m_{23}^2,\tag10\\
C'_2(\boldkey j)=-(m_{13}+j_1j_2)m_{23};\\
2)\quad m_{13}=j_2m_{13}^*,\quad m_{23}=j_1j_2j_3m_{23}^*,\\
C_2(\boldkey
j)=m_{23}^2+j_1^2j_3^2m_{13}(m_{13}+2j_2),\tag11\\
C'_2(\boldkey j)=-(m_{13}+j_2)m_{23};\\
3)\quad m_{13}=j_1j_2j_3m_{13}^*,\quad m_{23}=j_2m_{23}^*,\\
C_2(\boldkey
j)=m_{13}(m_{13}+2j_1j_2j_3)+j_1^2j_3^2m_{23}^2,\tag12\\
C'_2(\boldkey j)=-(m_{13}+j_1j_2j_3)m_{23}.\endgather$$
Considering (7), (8) only for operator $C_2(\boldkey j)$, the
following variants are admissible:
$$\gather
4)\quad m_{13}=j_1j_2j_3m_{13}^*,\quad m_{23}=m_{23}^*,\\
C_2(\boldkey
j)=m_{13}(m_{13}+2j_1j_2j_3)+j_1^2j_2^2j_3^2m_{23}^2,\tag13\\
C'_2(\boldkey j)=-j_2(m_{13}+j_1j_2j_3)m_{23};\\
5)\quad m_{13}=m_{13}^*,\quad m_{23}=j_1j_2j_3m_{23}^*,\\
C_2(\boldkey
j)=m_{23}^2+j_1^2j_2^2j_3^2m_{13}(m_{13}+2),\tag14\\
C'_2(\boldkey j)=-j_2(m_{13}+1)m_{23}\endgather$$
and other variants of
transformations of components $m^*_{13}$, $m^*_{23}$, which include
not all parameters $\boldkey j$, up to variant $m_{13}=m^*_{13}$,
$m_{23}=m^*_{23}$.

Considering contraction $\boldkey j={\boldsymbol\iota}$, we see
that general nondegenerate (with non-zero eigenvalues of both Casimir
operators) representations of contracted algebra
$so\,(4;{\boldsymbol\iota})$ come out only in the case of
transformations ``2'' and ``3''. In the case of transformations
``1'' we get $C_2({\boldsymbol\iota})=0$, in the case of
transformations ``4'', ``5'' -- $C'_2({\boldsymbol\iota})=0$.
Under other transformations both Casimir operators have zero
spectrum.

Let us consider variant ``3''. In this case components of scheme
``1'' satisfy inequalities
$${m_{13}\over j_1j_3}\geq|m_{23}|,\quad
{m_{13}\over j_1j_3}\geq{m_{12}\over j_3}\geq|m_{23}|,\quad
{m_{12}\over j_2}\geq|m_{11}|,
\tag15
$$
interpreted for imaginary and dual values of parameters $\boldkey j$
according to the rules represented in [10]. Using (12) and
rules of transformations for generators $X_{01}=j_1X^*_{01}$,
$X_{02}=j_1j_2X^*_{02}$, $X_{03}=j_1j_2j_3X^*_{03}$, we find operators
of irreducible representations of algebra $so\,(4;\boldkey j)$:
$$
\gather
X_{01}|m\rangle=im_{11}\beta|m\rangle-{1\over j_2j_3}
\alpha\sqrt{m_{12}(m_{12}^2-j_2^2m_{11}^2)}|m_{12}-j_2j_3\rangle+\\
+{1\over j_2j_3}\alpha(m_{12}+j_2j_3)\sqrt{(m_{12}+j_2j_3)^2-j_2^2m_{11}^2}
|m_{12}+j_2j_3\rangle,
\\
X_{02}|m\rangle={i\over2}\beta
[\sqrt{(m_{12}-j_2m_{11})(m_{12}+j_2m_{11}+j_2j_3)}|m_{11}+j_3\rangle+\\
+\sqrt{(m_{12}-j_2m_{11})(m_{12}-j_2m_{11}+j_2j_3)}|m_{11}-j_3\rangle]-\\
-{1\over2j_3}\alpha(m_{12})
\biggl[\biggr.
\sqrt{(m_{12}-j_2m_{11})(m_{12}-j_2m_{11}-j_2j_3)}\left|\matrix
\format\l\\ m_{12}-j_2j_3\\m_{11}+j_2\endmatrix\right>-\\
-\sqrt{(m_{12}+j_2m_{11})(m_{12}+j_2m_{11}-j_2j_3)}\left|\matrix
\format\l\\
m_{12}-j_2j_3\\m_{11}-j_3\endmatrix\right>
\biggl.\biggr]
-\\
-{1\over2j_3}\alpha(m_{12}+j_2j_3)\times\tag16\\
\times
\biggl[\biggr.
\sqrt{(m_{12}+j_2m_{11}+j_2j_3)(m_{12}+j_2m_{11}+2j_2j_3)}\left|\matrix
\format\l\\ m_{12}+j_2j_3\\m_{11}+j_3\endmatrix\right>-\\
-\sqrt{(m_{12}-j_2m_{11}+j_2j_3)(m_{12}-j_2m_{11}+2j_2j_3)}\left|\matrix
\format\l\\ m_{12}+j_2j_3\\m_{11}-j_3\endmatrix\right>
\biggl.\biggr]
,\\
X_{03}|m\rangle=j_3{1\over2}\beta
[\sqrt{(m_{12}-j_2m_{11})(m_{12}+j_2m_{11}+j_2j_3)}|m_{11}+\\+j_3\rangle
-\sqrt{(m_{12}+j_2m_{11})(m_{12}-j_2m_{11}+j_2j_3)}|m_{11}-j_3\rangle]+
\endgather
$$
$$
\gather
+{i\over2}\alpha(m_{12})
\biggl[\biggr.
\sqrt{(m_{12}-j_2m_{11})(m_{12}-j_2m_{11}-j_2j_3)}\left|\matrix
\format\l\\ m_{12}-j_2j_3\\m_{11}+j_3\endmatrix\right>+\\
+\sqrt{(m_{12}+j_2m_{11})(m_{12}+j_2m_{11}-j_2j_3)}\left|\matrix
\format\l\\
m_{12}-j_2j_3\\m_{11}-j_3\endmatrix\right>
\biggl.\biggr]+\\
+{i\over2}\alpha(m_{12}+j_2j_3)\times\\
\times
\biggl[\biggr.
\sqrt{(m_{12}+j_2m_{11}+j_2j_3)(m_{12}+j_2m_{11}+2j_2j_3)}\left|\matrix
\format\l\\ m_{12}+j_2j_3\\m_{11}+j_3\endmatrix\right>+\\
+\sqrt{(m_{12}-j_2m_{11}+j_2j_3)(m_{12}-j_2m_{11}+2j_2j_3)}\left|\matrix
\format\l\\ m_{12}+j_2j_3\\m_{11}-j_3\endmatrix\right>
\biggl.\biggr],\\
\alpha(m_{12})=\left\{{[(m_{13}+j_1j_2j_3)^2-j_1^2m_{12}^2]
(m_{12}^2-j_3^2m_{23}^2)\over
m_{12}^2 (4 m_{12}^2-j_2^2j_3^2)}\right\}^{1\over2},\\
\beta={(m_{13}+j_1j_2j_3)m_{23}\over
m_{12}(m_{12}+j_2j_3)}.
\endgather
$$

Here we presented all generators of algebra $so\,(4;\boldkey j)$ though,
as it has
been noticed, it is sufficient to give only $X_{01}$.

Initial finite-dimensional  irreducible representation of algebra
 $so\,(4)$ is Hermitean. The representation (16) of
algebra $so\,(4;\boldkey j)$ is irreducible, but, in general
non-Hermitean. To obtain Hermitean representation, it is necessary to
impose on operators (16) the requirement of Hermiticity:
$X_{\mu\nu}^\dagger=-X_{\mu\nu}$. It is difficult to find the
restrictions implied by this requirement on the components of
Gel'fand-Zetlin schemes. Therefore requirement of Hermiticity has to
be checked up in any particular case for concrete values of parameters
$\boldkey j$.

Considering variant (16), determined by (11), it turns out that
components of scheme (6) satisfy inequalities
$$m_{13}\geq{|m_{23}|\over j_1j_3},\quad
m_{13}\geq{m_{12}\over j_3}\geq{|m_{23}|\over j_1j_3},\quad
{m_{12}\over j_2}\geq|m_{11}|,\tag17$$
operators $X_{12}$, $X_{13}$, $X_{23}$ are described by (2), where
indices of parameters and generators have to be increased by one, and
operators $X_{0k}$ $(k=1.2.3)$ are given by (16), where functions
$\beta$ and $\alpha(m_{12})$ are substituted by functions
$\widetilde{\beta}$, $\widetilde{\alpha}(m_{12})$, which are as
follows:
$$\gather\widetilde{\alpha}(m_{12})=\left\{{[j_3^2(m_{13}+j_2)^2-m_{12}^2]
(j_1^2m_{12}^2-m_{23}^2)\over
m_{12}^2(4m_{12}^2-j_2^2j_3^2)}\right\}^{1\over2},\\
\widetilde{\beta}=
{(m_{13}+j_2)m_{23}\over
m_{12}(m_{12}+j_2j_3)}.
\tag18
\endgather
$$
\bigskip

{\bf 3. Contractions of representations
 of algebra $so\,(4;\boldkey j)$}
\medskip

Let us consider representations of algebra
$so\,(4;\iota_1,j_2,j_3)\equiv iso\,(3;j_2,\mathbreak j_3)=\{X_{0k}\}$
$\notni$ $so\,(3;j_2,j_3)$.  Let components $m_{13}$, $m_{23}$ are
transformed according to (12), i.e.
$k\equiv m_{13}=\iota_1j_2j_3m^*_{13}$, $m_{23}=j_2m^*_{23}$.
Operators of representation are described by (16) where
$$
\alpha(m_{12})=k\left\{{m_{12}^2-j_3^2m_{13}^2\over
m_{12}^2(4m_{12}^2-j_2^2j_3^2)}\right\}^{1\over2},\quad
\beta={km_{23}\over m_{12}(m_{12}+j_2j_3)}.
\tag19
$$
>From inequalities (15) for $j_2=j_3=1$ we find
$0\leq|m_{23}|<\infty$, $m_{12}\geq\mathbreak\geq|m_{23}|$,
$|m_{11}|\leq m_{12}$, where $m_{11}$, $m_{12}$, $m_{23}$ $\in\Bbb Z$;
$k\in\Bbb R$ (the latter -- from requirement of Hermiticity for
$X_{01}$). Spectrum of Casimir operators comes out of (12):
$C_2(\iota_1)=k^2$, $C'_2(\iota_1)=-km_{23}$.

If components are transformed according to (11), i.e.
$m_{13}=j_2m^*_{13}$, $s\equiv m_{23}=\iota_1j_2j_3m^*_{23}$ then
$\alpha$, $\beta$ are substituted for
$$
\widetilde{\alpha}(m_{12})=is\left\{{j_3^2(m_{13}+j_2)^2-m_{12}^2\over
m_{12}^2(4m_{12}^2-j_2^2j_3^2)}\right\}^{1\over2},\quad
\widetilde{\beta}={s(m_{13}+j_2)\over
m_{12}(m_{12}+j_2j_3)}.
\tag20
$$
Inequalities (17) for $j_2=j_3=1$
determine $m_{13}\geq m_{12}\geq0$, $|m_{11}|\leq m_{12}$,
$m_{11},m_{12},m_{13}\in\Bbb Z$, $s\in\Bbb R$. Spectrum of Casimir
operators comes out of (11):  $C_2(\iota_1)=s^2$,
$C'_2(\iota_1)=-sm_{13}$.

For algebra $so\,(4;j_1,j_2,\iota_3)=T_3$ $\notni$ $so\,(3;j_1,j_2)$,
where $T_3=\{X_{03},\mathbreak X_{13},X_{23}\}$ under transformation
(12), i.e.  $k\equiv m_{13}=j_1j_2\iota_3m^*_{13}$,
$m_{23}=\mathbreak=j_2m^*_{23}$, $p\equiv m_{12}=j_2\iota_3m^*_{12}$,
$q\equiv m_{11}=\iota_3m^*_{11}$, relations (16) determine operators
of irreducible representations:
$$
\gathered X_{01}|m\rangle=i{kqm_{23}\over p^2}|m\rangle+\\
+{1\over2p}\sqrt{(p^2-j^2_2q^2)(k^2-j_1^2p^2)}\left\{\right.2|m\rangle'_p+{p\over
p^2-j_2^2q^2}|m\rangle-\\
-{k^2\over p(k^2-j_1^2p^2)}|m\rangle\left.\right\},\quad
X_{03}|m\rangle=i\sqrt{k^2-j_1^2p^2}|m\rangle,\\
X_{02}|m\rangle=i{km_{23}\over p^2}\sqrt{p^2-j^2_2q^2}|m\rangle-\\
-{1\over2p}\sqrt{k^2-j_1^2p^2}\left\{\right.2p|m\rangle'_q+
2j^2_2q|m\rangle'_p-\\
-j_2^2{qk^2\over p(k^2-j_1^2p^2)}|m\rangle\left.\right\},\quad
|m\rangle=\left|\matrix
k&&m_{23}\\
&p&\\
&q&\endmatrix\right>.
\endgathered
\tag21
$$
The rest operators are given by (4) with obvious modifications.
Spectrum of Casimir operators is $C_2(\iota_3)=k^2$,
$C'_2(\iota_3)=-km_{23}$.

Inequalities (15) for $j_1=j_2=1$ imply $0\leq|m_{23}|<\infty$,
$k\geq p\geq0$, $|q|\leq p$, $m_{23}\in\Bbb Z$, $k,p,q\in\Bbb R$.

For transformation of components (11), i.e. $m_{13}=j_2m^*_{13}$,
$s\equiv\mathbreak\equiv m_{23}=j_1j_2\iota_3m^*_{23}$, $p\equiv
m_{12}=j_2\iota_3m^*_{12}$, $q\equiv m_{11}=\iota_3m^*_{13}$,
irreducible representation of algebra $so\,(4;j_1,j_2,\iota_3)$  is
described by operators
$$
\gather
X_{01}|\widetilde{m}\rangle=i{sq(m_{13}+j_2)\over
p^2}|\widetilde{m}\rangle
+{i\over2p}\sqrt{(p^2-j^2_2q^2)(j_1^2p^2-s^2)}\times\\
\times\left\{2|\widetilde{m}
\rangle'_p
+{s^2\over p(j_1^2p^2-s^2)}|\widetilde{m}\rangle+
{p\over p^2-j_2^2q^2}|\widetilde{m}\rangle\right\}, \\
X_{02}|\widetilde{m}\rangle=i{s(m_{13}+j_2)\over
p^2}\sqrt{p^2-j^2_2q^2}|\widetilde{m}\rangle- \tag22
\endgather
$$
$$
\gathered
-{i\over2p}\sqrt{j_1^2p^2-s^2}\left\{\right.2p|\widetilde{m}\rangle'_q+
2j^2_2q|\widetilde{m}\rangle'_p+j_2^2{s^2q\over
p(j_1^2p^2-s^2)}|\widetilde{m}\rangle\left.\right\},\\
X_{03}|\widetilde{m}\rangle=-\sqrt{j_1^2p^2-s^2}|\widetilde{m}\rangle,\quad
|\widetilde{m}\rangle=\left|\matrix
m_{13}&&s\\
&p&\\
&q&\endmatrix\right>.
\endgathered
$$
Components of scheme $|\,\widetilde{m}\rangle$ satisfy inequalities
implied by (11) for $j_1=\mathbreak=j_2=1$: $m_{13}\geq0$,
$p\geq|s|$, $|q|\leq p$, $m_{13}\in\Bbb Z$, $p,q,s\in\Bbb R$.
Spectrum of Casimir operators is as follows: $C_2(\iota_3)=s^2$,
$C'_2(\iota_3)=-s\,(m_{13}+j_2)$. It follows from (4) and (22)
that generators $X_{13}$, $X_{23}$,$X_{03}$ $\in T_3$ are diagonal in
continuous basis $|\,\widetilde{m}\rangle$.

Rejecting to fix coordinate axes, we notice that algebra
$so\,(4;\iota_1,1,1)$ is isomorphic to algebra $so\,(4;1,1,\iota_3)$,
and both these algebras are isomorphic to inhomogeneous algebra
$iso\,(3)$.  Isomorphism can be established by putting generator
$X_{\mu\nu}$, $\mu<\nu$ in correspondence with generator
$X_{3-\nu,3-\mu}$ of another algebra. Then operators (16) and (19)
determine irreducible representation of algebra $iso\,(3)$ in discrete
basis corresponding to the chain of subalgebras $iso\,(3)\supset
so\,(3)\supset so\,(2)$, and operators (21) and (4) describe the
same representation in continuous basis, corresponding to the chain
$iso\,(3)\supset so\,(3;1,\iota_3)\supset so\,(2;\iota_3)$. The same
assertion is valid for another variant of transition from
representation of algebra $so\,(4)$ to representations of algebra
$so\,(4;\boldkey j)$, which brings to (16), (20) and (4),
(22).

It is worth of noticing that contractions of representations give
another way of constructing the irreducible representations of
algebras (gro\-ups) with the structure of semidirect sum (product).

Operators of irreducible representation of algebra
$so\,(4;j_1,\iota_2,j_3)$ come out of (16) for $j_2=\iota_2$ and can
be written as follows:
$$
\gathered
X_{01}|m\rangle=i{ksm_{11}\over
p^2}|{m}\rangle+f(k,p,s)\left(2|m\rangle '_p+j_3^2{s^2\over
p(p^2-j_3^2s^2)}|m\rangle -\right.\\
-j_1^2{p^2\over k^2-j^2_1p^2}|m\rangle\left.\right),\\
X_{02}|{m}\rangle=i{ks\over
2p}(|m_{11}+j_3\rangle+|m_{11}-j_3\rangle)-
\endgathered
$$
$$
\gathered
-{1\over j_3}f(k,p,s)(|m_{11}+j_3\rangle-|m_{11}-j_3\rangle),\\
X_{03}|{m}\rangle=i{ks\over2p}(|m_{11}-j_3\rangle-|m_{11}-j_3\rangle)
+if(k,p,s)\times\\
\times(|m_{11}+j_3\rangle+|m_{11}-j_3\rangle),\\
f(k,s,p)={1\over2p}\sqrt{(k^2-j_1^2p^2)(p^2-j_3^2s^2)},\\
|{m}\rangle=
\left|\matrix
k&&s\\
&p&\\
&m_{11}&\endmatrix\right>.
\endgathered
\tag23
$$
Components of scheme $|\,m\rangle$ satisfy
inequalities: $k\geq|s|$, $-\infty<s<\infty$, $k\geq p\geq|s|$,
$-\infty<m_{11}<\infty$, $k,s,p\in\Bbb R$, $m_{11}\in\Bbb Z$, if
$j_1=j_3=\mathbreak=1$. Spectrum of Casimir operators is as follows:
$C_2(\iota_2)=k^2+j_1^2j_3^2s^2$, $C'_2(\iota_2)=-ks$.

For algebra $so\,(4;\iota_1,\iota_2,j_3)$ operators of irreducible
representation are given by (16) for $j_1=\iota_1$, $j_2=\iota_2$,
i.e.
$$
\gathered
X_{01}|m\rangle=i{ksm_{11}\over
p^2}|{m}\rangle+{k\over2p}\sqrt{p^2-j_3^2s^2}\times\\
\times\left(2|m\rangle'_p+j^2_3{s^2\over
p(p^2-j_3^2s^2)}|m\rangle\right),\\
X_{02}|{m}\rangle=i{ks\over
2p}(|m_{11}+j_3\rangle+|m_{11}-j_3\rangle)-\\
-{1\over j_3}{k\over2p}\sqrt{p^2-j^2_3s^2}
(|m_{11}+j_3\rangle-|m_{11}-j_3\rangle),\\
X_{03}|{m}\rangle=i{ks\over2p}(|m_{11}+j_3\rangle-|m_{11}-j_3\rangle)+\\
+{ik\over2p}\sqrt{p^2-j^2_3s^2}
(|m_{11}+j_3\rangle+|m_{11}-j_3\rangle).
\endgathered
\tag24
$$
Components of scheme $|\,m\rangle$ for $j_3=1$ satisfy inequalities:
$k\geq0$, $-\infty<\mathbreak<s<\infty$, $p\geq|s|$,
$-\infty<m_{11}<\infty$, $k,s,p\in\Bbb R$, $m_{11}\in\Bbb Z$. Spectrum
of Casimir operators is as follows: $C_2(\iota_1,\iota_2)=k^2$,
$C'_2(\iota_1,\iota_2)=-ks$.

Irreducible representation of algebra $so\,(4;j_1,\iota_2,\iota_3)$ is
given by
$$
\gathered
X_{01}|m\rangle=i{ksq\over
p}|{m}\rangle+{1\over2}\sqrt{k^2-j_1^2p^2}\times\\
\times\left(2|m\rangle'_p-j^2_1{p^2\over
(k^2-j_1^2p^2)}|m\rangle\right),\\
X_{02}|{m}\rangle=i{ks\over
p}|m\rangle-
\sqrt{k^2-j^2_1p^2}
|m\rangle'_q,\\
X_{03}|{m}\rangle=i\sqrt{k^2-j_1^2p^2}|m\rangle,\quad |m\rangle=
\left|\matrix
k&&s\\
&p&\\
&q&\endmatrix\right>.
\endgathered
\tag25
$$
which come out of (16) for $j_2=\iota_2$, $j_3=\iota_3$. For $j_1=1$
the components of scheme $|\,m\rangle$ satisfy inequalities:
$k\geq0$, $-\infty<s<\infty$, $k\geq p\geq0$, $-\infty<q<\infty$,
$k,s,p,q\in\Bbb R$. Spectrum of Casimir operators are the same as in
(24).

Algebra $so\,(4;\iota_1,\iota_2,j_3)=\widetilde{T}_5$ $\notni$
$so\,(2;j_3)$, where $so\,(2;j_3)=\{X_{23}\}$, is isomorphic to
algebra $so\,(4;j_1,\iota_2,\iota_3)=T'_5$ $\notni$ $so\,(2;j_1)$,
$so\,(2;j_1)=\{X_{01}\}$, and they both are isomorphic to algebra
$a=T_5$ $\notni$ $K$, where $T_5$ is nilpotent radical, and $K$ is
one-dimensional component subalgebra. Therefore (24) determine
irreducible representation of algebra $a$ in basis, corresponding to
the chain of subalgebras $so\,(4;\iota_1,\iota_2,j_3)$ $\supset$
$so\,(3;\iota_2,j_3)$ $supset$ $so\,(2;j_3)$, where
$so\,(2;j_3)=\{X_{23}\}$ is compact subalgebra with discrete
eigenvalues $m_{11}$, and (25) describe the same representation of
algebra $a$ in continuous basis, determined by the chain
$so\,(4;j_1,\iota_2,\iota_3)$ $\supset$ $so\,(3;\iota_2,\iota_3)$
$\supset$ $so\,(2;\iota_3)$; where $(so\,(2;\iota_3)=\{X_{23}\})$ is
already noncompact generator with continuous eigenvalues $q$.

For $j_1=\iota_1$, $j_3=\iota_3$ formulas (16) give irreducible
representation of algebra $so\,(4;\iota_1,j_2,\iota_3)$:
$$
\gathered
X_{01}|m\rangle=i{kqm_{23}\over
p^2}|{m}\rangle+{k\over2p}\sqrt{p^2-j_2^2q^2}\times\\
\times\left(2|m\rangle'_p+j^2_2{q^2\over
p(p^2-j_2^2q^2)}|m\rangle\right),
\endgathered
$$
$$
\gathered
X_{02}|{m}\rangle=i{km_{23}\over p^2}\sqrt{p^2-j^2_2q^2}|m\rangle-\\
-{k\over2p}\left(2j^2_2q|m\rangle'_p+2p|m\rangle'_q-j^2_2{q\over
p}|m\rangle\right), \\
X_{03}|{m}\rangle=ik|m\rangle,\quad |m\rangle=
\left|\matrix k&&m_{23}\\ &p&\\
&q&\endmatrix\right>.
\endgathered
\tag26
$$
Components of scheme $|\,m\rangle$ for $j_2=1$ satisfy following
inequalities: $k\geq0$, $-\infty<m_{23}<\infty$, $p\geq0$, $|q|\leq
p$, $k,p,q\in\Bbb R$, $m_{23}\in\Bbb Z$. Spectrum of Casimir operators
are as follows: $C_2(\iota_1,\iota_3)=k^2$,
$C'_2(\iota_1,\iota_3)=-km_{23}$. Three-dimensional contraction
$\boldkey j={\boldsymbol\iota}$ turns (16) into irreducible
representation of maximally contracted algebra
$so\,(4;{\boldsymbol\iota})$, described by operators
$$
\gathered
X_{01}|m\rangle=i{ksq\over
p^2}|{m}\rangle+k|m\rangle'_p\quad
X_{03}|{m}\rangle=ik|m\rangle,\\
X_{02}|{m}\rangle=i{ks\over p}|m\rangle-
k|m\rangle'_q,\quad
|m\rangle=
\left|\matrix k&&s\\ &p&\\ &q&\endmatrix\right>.
\endgathered
\tag27
$$
with eigenvalues of Casimir operators:
$C_2({\boldsymbol\iota})=k^2$, $C'_2({\boldsymbol\iota})=-ks$.
Components of scheme $|\,m\rangle$ are real, continuous and satisfy
following inequalities: $k\geq0$, $-\infty<s<\infty$, $p\geq0$,
$-\infty<q<\infty$.
\bigskip

{\bf 4. Algebra $so\,(n;\boldkey j)$}
\medskip

In [10] we have discussed in detail possible variants of
transformations of irreducible representation under transition from
algebra $so\,(4)$ to algebra $so\,(4;\boldkey j)$.  For orthogonal
algebras of arbitrary dimension we shall not consider all possible
variants, but dwell on the (basic) variant, in which number of
parameters $\boldkey j$, on which components some row of
Gel'fand-Zetlin schemes, diminishes with increasing of component
number in this row. The transformation of components under transition
from algebra $so\,(n)$ to $so\,(n;\boldkey j)$ can be found from the
rule of transformation for Casimir operators. Because orthogonal
algebras of even and odd dimensions have different sets of Casimir
operators, we shall discuss these cases separately.

Algebra $so\,(2k+2;\boldkey j)$, $\boldkey j=(j_1,\dots,j_{2k+1})$, is
characterized by a set of $k+1$ invariant operators [1]
$$ \gathered
C_{2p}(\boldkey j)
=\prod_{s=1}^{p-1}j_s^{2s}j_{2(k+1)-s}^{2s}\prod_{l-p}^{2(k+1)-p}
j_l^{2p}C^*_{2p}(\to),\quad p=1,2,\dots,k,\\
C'_{k+1}(\boldkey
j)=j_{k+1}^{k+1}\prod_{l=1}^{k}j_l^lj_{2(k+1)-l}^lC^{*'}_{k+1}(\to).
\endgathered
\tag28
$$
Gel'fand-Zetlin scheme for
algebra $so\,(2k+2)$ is as follows:
$$
|m\rangle^*=\left|\matrix
m^*_{1,2k+1}\quad m^*_{2,2k+1}& \dots& m^*_{k,2k+1}\quad
m^*_{k+1,2k+1}\\
\vspace{2pt}
m^*_{1,2k}& \dots& m^*_{k,2k}\\
\vspace{2pt}
\ \ \ m^*_{1,2k-1}&\dots& m^*_{k,2k-1}\\
\vspace{2pt}
\phantom{m^*_{1,2k+1}\quad} m^*_{1,2k-2}& \dots&
m^*_{k-1,2k-2}\phantom{\quad
m^*_{k+1,2k+1}}\\
\vspace{2pt}
\phantom{m^*_{1,2k+1}\quad} m^*_{1,2k-3}& \dots&
m^*_{k-1,2k-3}\phantom{\quad
m^*_{k+1,2k+1}}\\
\vspace{2pt}
\hdotsfor3\\
\vspace{2pt}
&m^*_{12}&\\
&m^*_{11}&\endmatrix\ \ \right>.
\tag29
$$
Irreducible representation as
well as spectrum of Casimir operators on this representation are
completely determined by components $m^*_{p,2k+1}$ of the upper row of
scheme (29). Component $m^*_{1,2k+1}$ enters spectrum of Casimir
operator $C^*_2$ quadratically, and due to this fact in the basic
variant it is transformed according to the rule
$m_{1,2k+1}=m^*_{1,2k+1}\prod\limits_{l=1}^{2k+1}j_l$. The
transformation of the component $m^*_{p,2k+1}$ coincides with
transformation of algebraic quantity
$\{C^*_{2p}/C^*_{2(p-1)}\}^{1/2}$:
$$
m_{p,2k+1}=m_{p,2k+1}^*\prod_{l=p}^{2(k+1)-p}j_l\equiv
m_{p,2k+1}^*J_{p,2k+1}.
\tag30
$$
Component $m^*_{k+1,2k+1}$ is transformed in
the same way as the ratio
$C_{k+1}^{*'}/\mathbreak/\{C^*_{2k}\}^{1/2}$, i.e.
$m_{k+1,2k+1}=j_{k+1}m^*_{k+1,2k+1}$. This relation comes out of
(30) for $p=k+1$ as well as relation for transformation of component
$m^*_{1,2k+1}$ that comes out of (30) for $p=1$. Thus, all components
of major weight (the upper row) are transformed according to (30).
Inequalities, which governed them in classical case, turns into $$
\gather
{m_{p,2k+1}\over J_{p,2k+1}}\geq{m_{p+1,2k+1}\over J_{p+1,2k+1}},\\
{m_{k,2k+1}\over J_{k,2k+1}}\geq{|m_{k+1,2k+1}|\over
J_{k+1,2k+1}},\tag31\\
p=1,2,\dots,k-1.
\endgather
$$

Similarly, transformation of components of the row with number $2k$
of scheme (29) is determined by the rules of transformations for
Casimir operators of subalgebra $so\,(2k+1;j_2,j_3,\dots,j_{2k+1})$
and given by (30), in which product of parameters $j_l$ starts with
$p+1$. In general the rule of transformation for components of scheme
(29) can be easily found and turns out to be as follows:
$$
\gathered
m_{p,2s+1}=J_{p,2s+1}m^*_{p,2s+1},\quad
J_{p,2s+1}=\prod_{l=p+2(k-s)}^{2(k+1)-p}j_l,\\
s=0,1,\dots,k,\quad p=1,2,\dots,s+1,\\
m_{p,2s}=J_{p,2s}m^*_{p,2s},\quad
J_{p,2s}=\prod_{l=p+2(k-s)+1}^{2(k+1)-p}j_l,\\
s=1,2,\dots,k,\quad p=1,2,\dots,s.
\endgathered
\tag32
$$
The transformed components are governed by inequalities
$$
\gathered
{m_{p,2s+1}\over J_{p,2s+1}}\geq{m_{p,2s}\over J_{p,2s}}
\geq{m_{p+1,2s+1}\over J_{p+1,2s+1}},\quad p=1,2,\dots,s-1,\\
{m_{s,2s+1}\over J_{s,2s+1}}\geq{m_{s,2s}\over J_{s,2s}}
\geq{|m_{s+1,2s+1}|\over J_{s+1,2s+1}},\\
{m_{p,2s}\over J_{p,2s}}\geq{m_{p,2s-1}\over
J_{p,2s-1}}\geq{m_{p+1,2s}\over J_{p+1,2s}},\quad p=1,2,\dots,s-1,\\
{m_{s,2s}\over J_{s,2s}}\geq{m_{s,2s-1}\over J_{s,2s-1}}
\geq-{m_{s,2s}\over J_{s,2s}},
\endgathered
\tag33
$$
which for dual
and imaginary values of parameters $\boldkey j$ are interpreted
according to the rules described in [10].

Action of the whole algebra $so\,(2k+2;\boldkey j)$ can be reproduced
by giving the action of generators $X_{2(k-s)+1,2(k-s+1)}$,
$s=1,2,\dots,k$,\linebreak$X_{2(k-s),2(k-s)+1}$,
$s=0,1,\dots,k-1$.  Transforming the expressions for these
generators, which can be found in [7], we obtain $$ \gather
X_{2(k-s)+1,2(k-s+1)}|m\rangle=\sum_{p=1}^{s}{1\over
J_{p,2s-1}}\{A(m_{p,2s-1})|m_{p,2s-1}+\\
+J_{p,2s-1}\rangle-A(m_{p,2s-1}-J_{p,2s-1})|m_{p,2s-1}-J_{p,2s-1}\rangle\},\\
X_{2(k-s),2(k-s)+1}|m\rangle=iC_{2s}|m\rangle+\sum_{p=1}^{s}{1\over
J_{p,2s}}\{B(m_{p,2s})|m_{p,2s}+\\
+J_{p,2s}\rangle-B(m_{p,2s}-J_{p,2s})|m_{p,2s}+J_{p,2s}\rangle\},\\
C_{2s}=\prod_{p=1}^{s}l_{p,2s-1}\prod_{p=1}^{s+1}l_{p,2s+1}
\prod_{p=1}^{s}l_{p,2s}^{-1}(l_{p,2s}-J_{p,2s})^{-1},\\
B(m_{p,2s})=\biggl\{\biggr.\prod_{r=1}^{p-1}(l_{r,2s-1}^2-l_{p,2s}^2
a_{r,p,s}^2)\prod_{r=p}^{s}(l_{r,2s-1}^2a_{r,p,s}^{-2}-l_{p,2s}^2)
\times\tag34\\
\times\prod_{r=1}^{p}(l_{r,2s+1}^2-l_{p,2s}^2b_{r,p,s}^2)
\prod_{r=p+1}^{s+1}(l_{r,2s+1}^2b_{r,p,s}^{-2}-l_{p,2s}^2)\biggl.\biggr\}
^{1\over2}\times\\
\times\biggl\{\biggr.l^2_{p,2s}(4l^2_{p,2s}-J_{p,2s}^2)\prod_{r=1}^{p-1}(l_{r,2s}^2-
l_{p,2s}^2c^2_{r,p,s})[(l_{r,2s}-J_{r,2s})^2-l_{p,2s}c_{r,p,s}^2]\times\\
\times\prod_{r=p+1}^{s}(l^2_{r,2s}c^{-2}_{r,p,s}-l^2_{p,2s})[(l_{r,2s}-
J_{r,2s})^2c^{-2}_{r,p,s}-l^2_{p,2s}]\biggl.\biggr\}^{-{1\over2}},\\
A(m_{p,2s-1})={1\over2}\biggl\{\biggr.\prod_{r=1}^{p-1}(l_{r,2s-2}-
l_{p,2s-1}a_{r,p,s-1/2}-J_{r,2s-2})\times\\
\times(l_{r,2s-2}
+l_{p,2s-1}a_{r,p,s-1/2})\prod_{r=p}^{s-1}(l_{p,2s-2}a^{-1}_{r,p,s-1/2} -
l_{p,2s-1}-
\endgather
$$
$$
\gather
-J_{p,2s-1})(l_{r,2s-2}a^{-1}_{r,p,s-1/2}+l_{p,2s-1})\times\\
\times\prod_{r=1}^{p}(l_{r,2s}-l_{p,2s-1}b_{r,p,s-1/2}-J_{r,2s})(l_{r,2s}+\\
+l_{p,2s-1}b_{r,p,s-1/2}\prod_{r=p+1}^{s}(l_{r,2s}b_{r,p,s-1/2}^{-1}-
l_{p,2s-1}-J_{p,2s-1})\times\\
\times(l_{r,2s}b_{r,p,s-1/2}^{-1}+l_{p,2s-1})\biggl.\biggr\}^{1\over2}\biggl\{
\biggr.\prod_{r=p+1}^{s}(l^2_{r,2s-1}c_{r,p,s-1/2}^{-2}-l_{p,2s-1}^2)\times\\
\times[l^2_{r,2s-1}-(l_{r,2s-1}+J_{p,2s-1})c_{r,p,s-1/2}^{2}]\prod_{r=p+1}^{s}
l^2_{r,2s-1}c_{r,p,s-1/2}^{-2}-\\
-l^2_{p,2s-1})[l^2_{r,2s-1}c_{r,p,s-1/2}^{-2}-(l_{p,2s-1}+J_{p,2s-1})^2]
\biggl.\biggr\}^{-{1\over2}},\\
a_{r,p,s}=J_{r,2s-1}/J_{p,2s},\quad b_{r,p,s}=J_{r,2s+1}/J_{p,2s},\\
c_{r,p,s}=J_{r,2s}/J_{p,2s},\quad
l_{p,2s}=m_{p,2s}+(s-p+1)J_{p,2s}.
\endgather
$$

For algebra $so\,(2k+1)$ Gel'fand-Zetlin scheme $|\,m^*\rangle$
coincides with (29) with deleted row with number $2k+1$. The upper
row, determining the components of major weight, now is the row with
number $2k$; its components satisfy inequalities $m^*_{1,2k}\geq
m^*_{2,2k}\geq\dots\geq m^*_{k,2k}\geq0$.  Under transition from
classical algebra $so\,(2k+1)$ to algebras $so\,(2k+1;\boldkey j)$,
$\boldkey j=(j_1,\dots,j_{2k})$ the components of scheme $|\,m\rangle$
are transformed as follows:
$$
\gather
m_{p,2s}=m_{p,2s}^*J_{p,2s},\quad
J_{p,2s}=\prod_{l=p+2(k-s)}^{2k+1-p}j_l,\\
m_{p,2s-1}=m_{p,2s-1}^*J_{p,2s-1},\quad
J_{p,2s-1}=\prod_{l=p+2(k-s)+1}^{2k+1-p}j_l,\tag35\\
s=1,2,\dots,k,\quad p=1,2,\dots,s.
\endgather
$$
Let us draw attention to the fact that
the lower limits in the product, determining $J_{p,2s}$, $J_{p,2s-1}$,
have changed in comparison with (32). This is due to the diminishing
of the number of parameters $\boldkey j$ by one in the case of algebra
$so\,(2k+1;\boldkey j)$ in comparison with algebra
$so\,(2(k+1);\boldkey j)$. Components of the upper row in scheme
$|\,m\rangle$ satisfy inequalities
$$
{m_{1,2k}\over J_{1,2k}}\geq{m_{2,2k}\over J_{2,2k}}
\geq\dots\geq{m_{k-1,2k}\over J_{k-1,2k}}\geq{m_{k,2k}\over
J_{k,2k}}\geq0,
\tag36
$$
and the other components are
governed by inequalities (33), which parameters $J_{p,2s}$,
$J_{p,2s-1}$ are defined according to (35). Operators of irreducible
representation of algebra $so\,(2k+1;\boldkey j)$ are given by (34)
with parameters from (35).

Operators (34) satisfy
commutation relations of algebra $so\,(n;\boldkey j)$.\linebreak This
can be checked up by straightforward calculations as well. The
irreducibility of representation is implied by consideration of action
of rising and lowering operators on vectors of major and minor weights
and non-zero result of this action for dual and imaginary values of
parameters $\boldkey j$. Though the initial representation of algebra
$so\,(n)$ is Hermitean, the representation (34), in general is not
such.  Requiring the fulfillment of condition
$X^\dagger_{\mu\nu}=-X_{\mu\nu}$, we find those values of transformed
components of Gel'fand-Zetlin scheme, for which representation (34)
will be Hermitean.

It is worth of noticing, at last, that for imaginary values of
parameters $\boldkey j$ the relations (34) give irreducible
representations of pseudoorthogonal algebras of different signature.

\medskip
The authors are grateful to Yu.A.~Danilov for helpful discussions.

\vskip1cm
\centerline{\bf References}

\medskip
\item{1.}
Gromov N.A., Moskaliuk S.S. Special orthogonal groups in Cayley--Klein
spaces // Hadronic Journal. -- 199\ \ . N \ \ \  -- P.

\medskip
\item{2.}
Gromov N.A., Moskaliuk S.S. Special unitary  groups in
Cayley--Klein spaces // Hadronic Journal. -- 199\ \ . N \ \
\  -- P.

\medskip
\item{3.}
Gel'fand I.M., Zetlin M.L. Finite-dimensional representations of\linebreak group
of unimodular matrices // Dokl. of Acad. Sci. USSR. Math. se\-ri\-es. 
-- 1950. -- \underbar{\bf 71}, N 5. -- P. 825--828.

\medskip
\item{4.}
Celegnini E., Tarlini M. Conractions of group representations. I //
Nuovo Cim. B. -- 1981. -- \underbar{\bf 61}, N 2. -- P.
265--277.

\medskip
\item{5.}
Celegnini E., Tarlini M. Conractions of group representations. II //
Nuovo Cim. B. -- 1981. -- \underbar{\bf 65}, N 1. -- P.
172--180.

\medskip
\item{6.}
Celegnini E., Tarlini M. Conractions of group representations. III //
Nuovo Cim. B. -- 1982. -- \underbar{\bf 68}, N 1. -- P.
133--141.

\medskip
\item{7.}
Barut A., Ronczka P. Theory of group representations. Moscow: Mir,
1980. -- Vol.1 -- 456 p.; Vol.2 -- 396 p.

\medskip
\item{8.}
Perelomov A.M., Popov V.S. Casimir operators for orthogonal and
symplectic groups // Nucl. Phys. -- 1966. -- \underbar{\bf 3},
N 6. -- P. 1127--1134.

\medskip
\item{9.}
Leznov A.N., Malkin I.A., Man'ko V.I. Canonical transformations and
theory of representations of Lie
 groups // Proc. of Phys. Inst. of Acad. Sci. USSR. --
1977. -- \underbar{\bf 96}. -- P. 27--71.

\medskip
\item{10.}
Gromov N.A., Moskaliuk S.S. Irreducible representations of
Cayley--Klein unitary algebras // Proceedings of International
Workshop\linebreak ``New Frontiers in Algebras and Groups'', Monteroduni--Haly,
August 1995. Vol.1. -- Hadronic Press, Palm Harbor, FL U.S.A., 1996.
-- P. 361--392.

\end